\documentclass[twocolumn,9pt]{article}

\usepackage{balance}
\usepackage{amsmath,amssymb,amsfonts}
\usepackage{algorithmic}
\usepackage{graphicx}
\usepackage{textcomp}
\usepackage{xcolor}
\usepackage{multirow}
\usepackage{csquotes}
\usepackage{comment}
\usepackage{float}
\usepackage[hyphens]{url}
\usepackage[hidelinks]{hyperref}
\usepackage{authblk}
\begin{document}

\title{First Experiments with a 5G-Connected Drone}

\author[1]{Raheeb Muzaffar}
\author[1]{Christian Raffelsberger}
\author[2]{Aymen Fakhreddine}
\author[3]{\\Jos\'{e} L\'{o}pez Luque}
\author[4]{Driton Emini}
\author[1,2]{Christian Bettstetter}
\affil[1]{Lakeside Labs GmbH, Klagenfurt, Austria}
\affil[2]{University of Klagenfurt, Klagenfurt, Austria}
\affil[3]{Magenta Telekom GmbH, Vienna, Austria}
\affil[4]{Deutsche Telekom AG, Bonn, Germany}

\date{}

\maketitle

\begin{abstract}
We perform experiments on the wireless communication between a drone flying at different heights and a commercial 5G base station. An Android-based tool deployed on a 5G test platform is used to record radio link parameters in the up- and downlink. In the downlink, measurements show a throughput of 600\,Mbit/s on average with peaks above 700\,Mbit/s. The uplink has a significantly lower throughput, comparable to 4G, with a few tens of Mbit/s.
\end{abstract}

\pagestyle{plain}

\sloppy
\section{Introduction}
\label{intro}
Drones need wireless connectivity to transfer commands, images and videos, sensor measurements, and other mission-oriented data~\cite{7463007,zeng2019accessing,mozaffari2018tutorial}. The requirements on the wireless technology to be used are very application-specific~\cite{andre14:commag}. Commonly used short-range unlicensed radio technologies\,---\,such as Wi-Fi, Bluetooth, and Zigbee\,---\,are not fully suited for some commercial drone applications due to demands in terms of radio coverage, throughput, latency, and scalability~(see \cite{yang2018telecom,7463007}). Existing fourth generation (4G) cellular networks offer wide-area coverage with data rates acceptable for certain drone systems but insufficient for advanced, autonomously organizing multi-drone systems that require high-rate real-time communication and computation offloading. The ongoing rollout of fifth generation~(5G) cellular networks is expected to improve the situation with new features such as enhanced mobile broadband, ultra-reliable low latency communication, network slicing, and mobile edge computing. In fact, standardization efforts are underway to incorporate drones into 5G networks~\cite{3GPP36777,8108590,9061157}.

\begin{figure}[t!]
\begin{center}
\scriptsize
\includegraphics[width=0.45\textwidth]{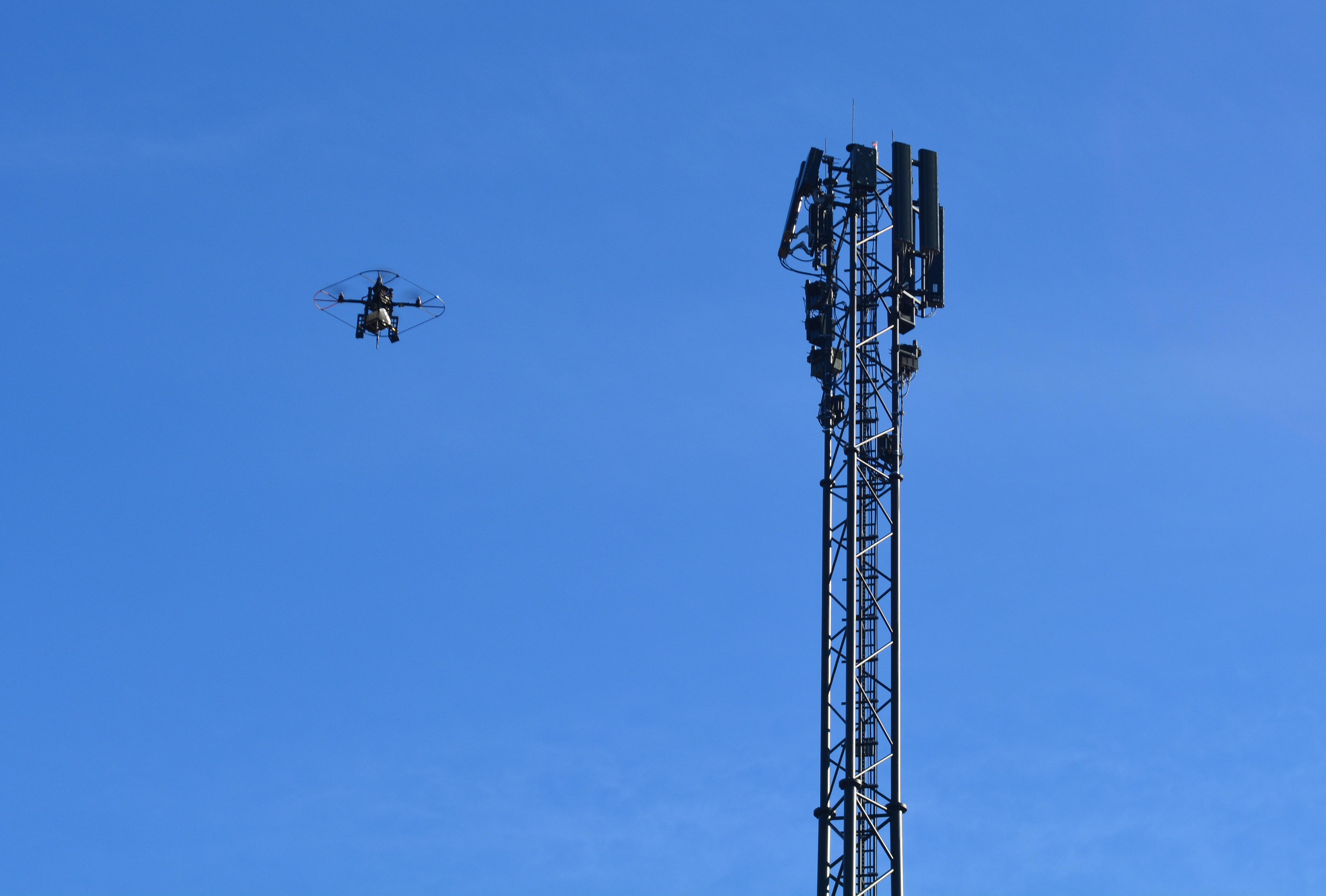}
\end{center}
\caption{Quadrocopter with 5G user equipment flying close to a 5G base station} 
\label{fig:Drone_Tower}
\end{figure}

This paper examines 5G connectivity for drones experimentally, taking into account the current state of 5G deployment. In simple terms, our key question is: What happens if we connect a flying drone to one of today's commercially operated 5G systems? Although cellular-connected drones have been investigated in the past few years (see, e.g., \cite{8692749,athanasiadou2019lte,becvar2017performance,8579209}), to our knowledge, this is the first peer-reviewed scientific paper with  experimental results on radio link quality and throughput between a drone and a 5G base station in commercial operation. The measurements are carried out using a custom-built measurement tool~\cite{raffelsberger19:cdmt} in one of the 5G new radio systems deployed by Magenta Telekom in Austria. To do so, we fly a quadrocopter carrying a 5G user equipment at different heights and different distances in the vicinity of a 5G base station with three sectors (Fig.~\ref{fig:Drone_Tower}). The downlink throughput is about $600$~Mbit/s on average with peaks above $700$~Mbit/s. However, frequent handovers to 4G are observed, causing high fluctuations of the overall throughput. As there are no other 5G base stations in the vicinity of our~experiments, handovers to 4G are inevitable. 
This setup reflects the actual situation during the transition phase from 4G to 5G. The results are not representative for a fully-deployed 5G network, but they are valuable as they provide basic insight and serve as a benchmark for further experiments to be conducted in a full 5G network.

The remainder of the paper is structured as follows. Section~\ref{section:related-work} covers related experimental work on cellular-connected drones and summarizes standardization efforts. Section~\ref{Setup} describes the experimental setup. Section~\ref{results} is the core part of the paper: it presents and discusses the experimental results for three different flights. Finally, Section~\ref{concl}~draws general conclusions and states open issues. 

\begin{figure}[t]
\begin{center}
\scriptsize
\includegraphics[width=0.45\textwidth]{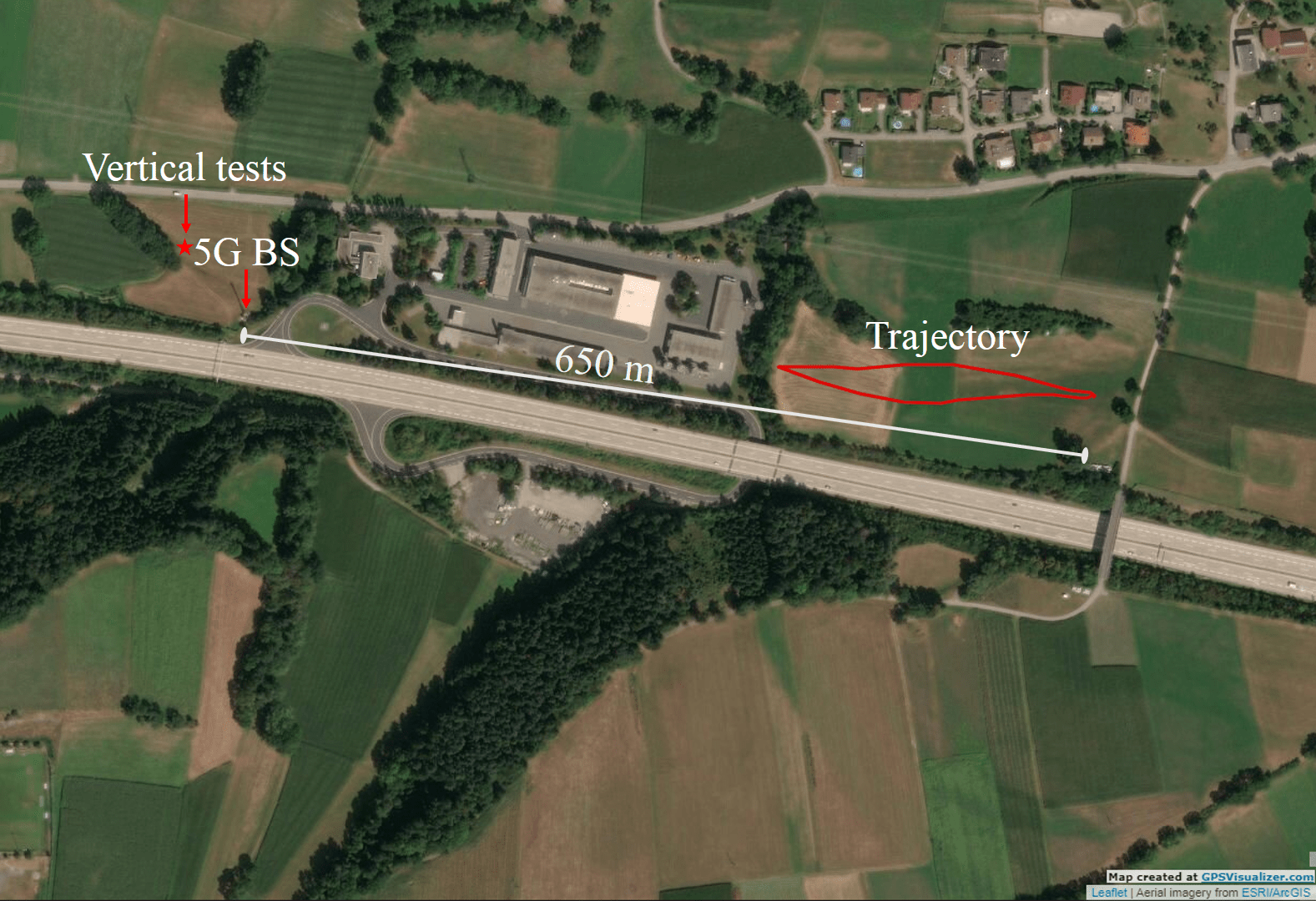}
\end{center}
\caption{Flight trajectory and location of vertical tests} 
\label{fig:Flight_trajectory}
\end{figure}

\section{Related Work}\label{section:related-work}
Previous experimental studies on cellular-connected drones study throughput, interference, and handovers in 4G networks (see, e.g., \cite{8287891,hayat19,fakhreddine19,nguyen2018ensure,athanasiadou2019lte,lin2019mobile}). An average throughput of a few tens of Mbit/s can be achieved both in uplink and downlink. One of the main issues is that drones experience line-of-sight~(LoS) links to  far distant base stations, which are normally ``invisible" to regular ground users. These LoS links cause interference in base stations and aerial drones~\cite{van2016lte,bertizzolo2020live} and lead to frequent and unnecessary handovers~\cite{fakhreddine19}. Interference degrades the throughput for both terrestrial and aerial users~\cite{hayat19,muruganathan2018overview}. A degraded performance implies that data transmission takes longer and hence requires more~resources~\cite{muruganathan2018overview}. 

The 3GPP~(3rd Generation Partnership Project) Release~15 addresses some of these issues. The technical specification defines the requirements for enhanced support of aerial vehicles~\cite{3GPP36777}. Data rates between $40$ and $60$~kbit/s are suggested for command-and-control signaling with a maximum one-way latency of $50$~ms to ensure proper operational control of the aerial vehicles. The uplink application data rate of $50$~Mbit/s is suggested with a latency similar to 4G terrestrial~users~\cite{3GPP36777}.

Potential enhancements to support connectivity for aerial devices includes reporting measurements (such as height estimation and radio link quality) from the aerial devices. Such information would be useful for base stations to identify a drone that might experience interference. Furthermore, flight paths can be reported to enable the network to plan for required resources~\cite{3GPP21915,muruganathan2018overview}.

\section{Experimental Setup}
\label{Setup}
The purpose of our study is to analyze 5G radio links for use with drones. The experiments are performed at a site in Feichtendorf, Austria, where Magenta operates a 5G base station (BS). We fly an Asctec Pelican quadrocopter in an open field adjacent to this BS (see Fig.~\ref{fig:Flight_trajectory}). Three types of flights are performed: a liftoff from the ground to a height of $150$~m (vertical flight $70$~m away from the BS) and two horizontal flights where the drone moves at a height of $30$~m and $100$~m above ground, respectively. The drone starts $650$~m away from the BS, flies $250$~m toward the BS with a speed of $3$~m/s, and returns to the starting point.

The BS uses 5G New Radio (NR) with 3GPP Release~15  operating at frequencies between $3.7$ and $3.8$~GHz using a $100$~MHz band. The system shares one band for downlink~(DL) and uplink~(UL) and assigns time slots for transmitting and receiving through time-division duplexing (TDD). The 5G BS has $64 \times 64$ massive MIMO (multiple-input multiple-output) antennas with beamforming capabilities, which can electrically change the tilt from $-2^\circ$ to $9^\circ$. The horizontal $3$~dB beamwidth is $105^\circ$ and the vertical one is~$6^\circ$. Besides the 5G NR, the site is also connected to the Magenta Telekom 4G network. It runs the 3GPP Release $15$ in the $800$~MHz and $1800$~MHz bands. These two bands can be combined with carrier aggregation  to provide higher data~rates.

\begin{figure}[h!]
\begin{center}
\scriptsize
\includegraphics[width=0.35\textwidth]{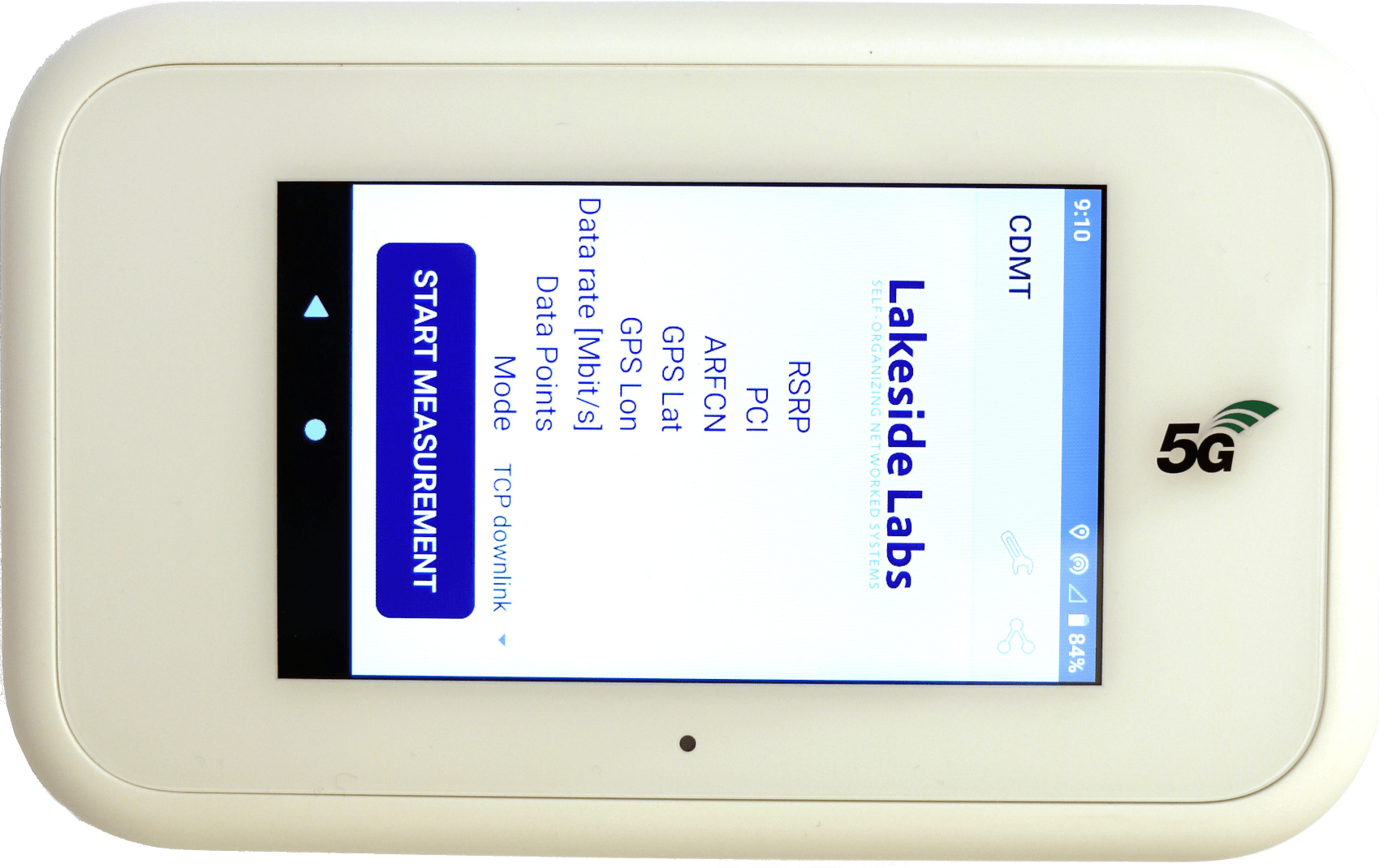}
\end{center}
\caption{Mobile test platform and application} 
\label{fig:5G-device}
\end{figure}

\begin{table*}[!htb]
\centering
\renewcommand{\arraystretch}{1.37}
\caption{Measurement results}
\begin{tabular}{|lc|r|r|r|r|c|c|}
\hline
\multirow{2}{*}{\bf{Experiment}} & \multirow{2}{*}{\bf{Link}} & \multicolumn{4}{c|}{\bf{Throughput}} & {\bf{Time}} &\multirow{2}{*}{\bf{Handovers}}\\ 
\cline{3-6}
&&  {maximum} & {mean~~~} & {stddev~~~} & {5G mean} & {\bf in 5G} & \\
\hline \hline
Liftoff & DL &  $742$\,Mbit/s & $345$\,Mbit/s & $244$\,Mbit/s & $387$\,Mbit/s & $67$\,\% & $1$ \\
 & UL  &  $64$\,Mbit/s & $44$\,Mbit/s & $8$\,Mbit/s & $39$\,Mbit/s & $93$\,\% & $2$ \\\hline
Horizontal flight at 30~m & DL &  $713$\,Mbit/s & $388$\,Mbit/s & $273$\,Mbit/s & $618$\,Mbit/s & $57$\,\% & $2$\\ 
& UL &  $51$\,Mbit/s & $46$\,Mbit/s & $2$\,Mbit/s & $46$\,Mbit/s & $100$\,\% & $0$ \\\hline
Horizontal flight at 100~m& DL &  $707$\,Mbit/s & $354$\,Mbit/s & $306$\,Mbit/s & $644$\,Mbit/s & $53$\,\% & $3$ \\ 
 & UL  & $67$\,Mbit/s & $47$\,Mbit/s & $8$\,Mbit/s & $42$\,Mbit/s & $66$\,\% & $5$ \\\hline
\end{tabular}
\label{tab:performance}
\end{table*}

The drone carries a Wistron NeWeb non-standalone mobile test platform based on the Qualcomm Snapdragon~X50 5G~modem (Fig.~\ref{fig:5G-device}). The platform supports 5G NR at sub-$6$~GHz frequency and uses $256$~QAM~(quadrature amplitude modulation) with \mbox{$4\times4$ MIMO} that features data rates up to $2.22$~Gbit/s~\cite{5G2019}. Furthermore, it supports 4G and has the ability to connect to both 4G and 5G networks simultaneously through E-UTRAN New Radio Dual Connectivity (ENDC). The Cellular Drone Measurement Tool (CDMT)~\cite{raffelsberger19:cdmt} is used to record the measurements. CDMT is based on an Android API~(application programming interface) and employs a client-server model, where the client is the application on the drone, and the server resides at the Lakeside Science and Technology Park in Klagenfurt, Austria. The server is accessible from the public Internet via a symmetrical $1$~Gbit/s connection. We refer to the communication of the data from (to) the server via 5G to (from) the drone as~DL~(UL).

\section{Experimental Results}
\label{results}

The communication of the 5G-connected drone is assessed in terms of the reference signal received power~(RSRP), signal-to-noise ratio~(SNR), throughput, 5G connectivity, and number of handovers at different flight heights. Table~\ref{tab:performance} summarizes some measurement results. They are based on data collected from a single drone flight for each setup.  Repetitions of the given setup led to similar qualitative~results.

\subsection{Liftoff}

\begin{figure}[h!]
\begin{center}
\centering 
\includegraphics[width=3.4in]{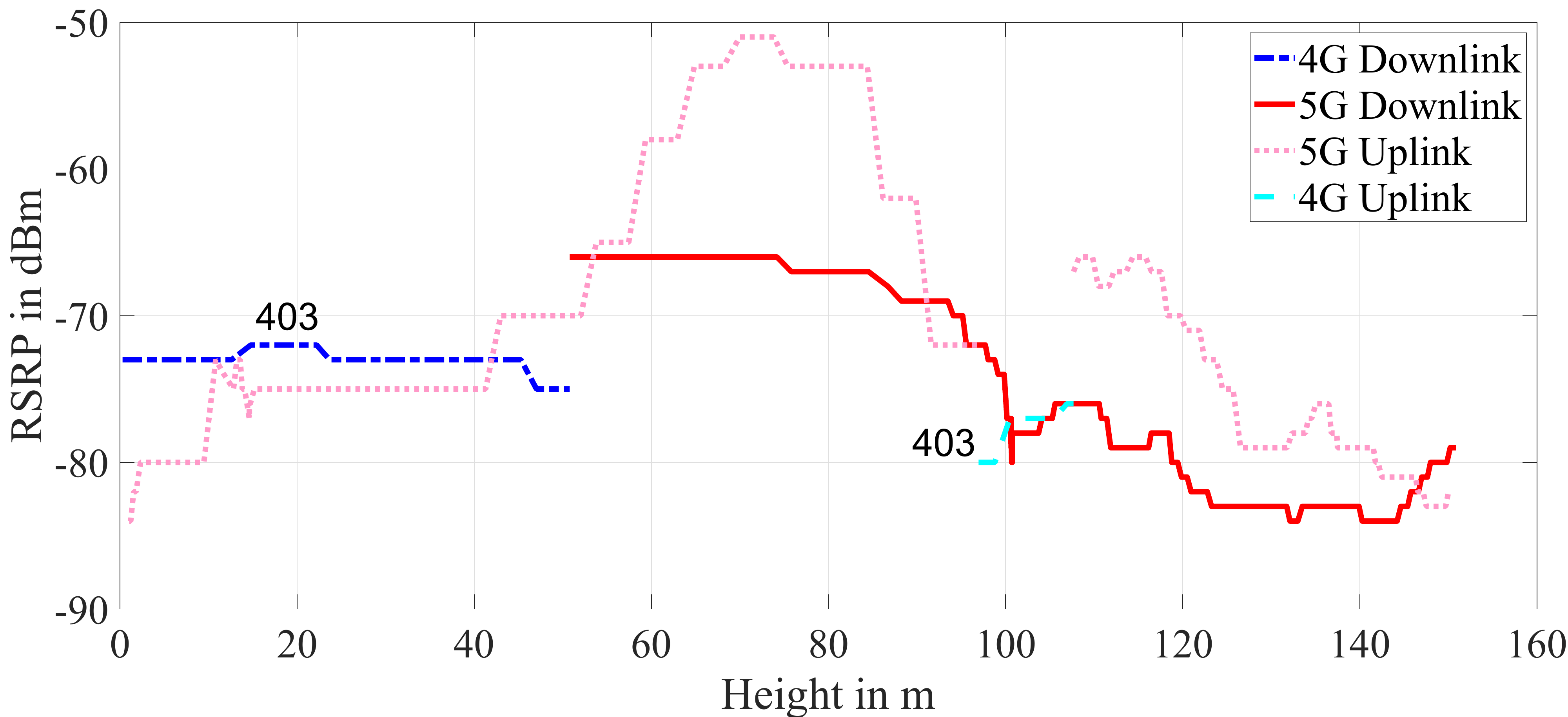} \\ 
\hspace{-0.3cm}(a) RSRP (Physical Cell ID~(PCI)~403) \\[0.3cm] 
\includegraphics[width=3.4in]{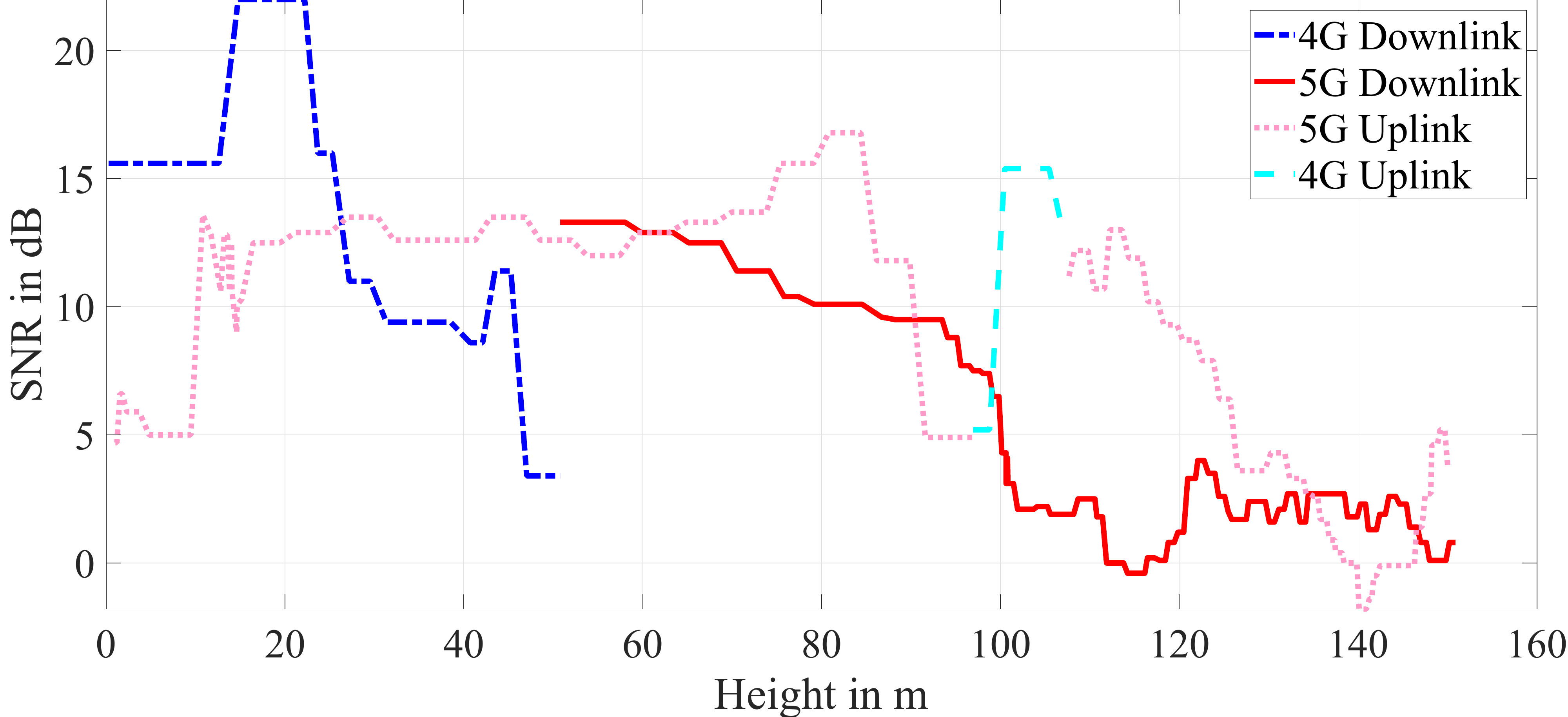} \\
\hspace{-0.3cm}(b) SNR \\[0.3cm] 
\includegraphics[width=3.4in]{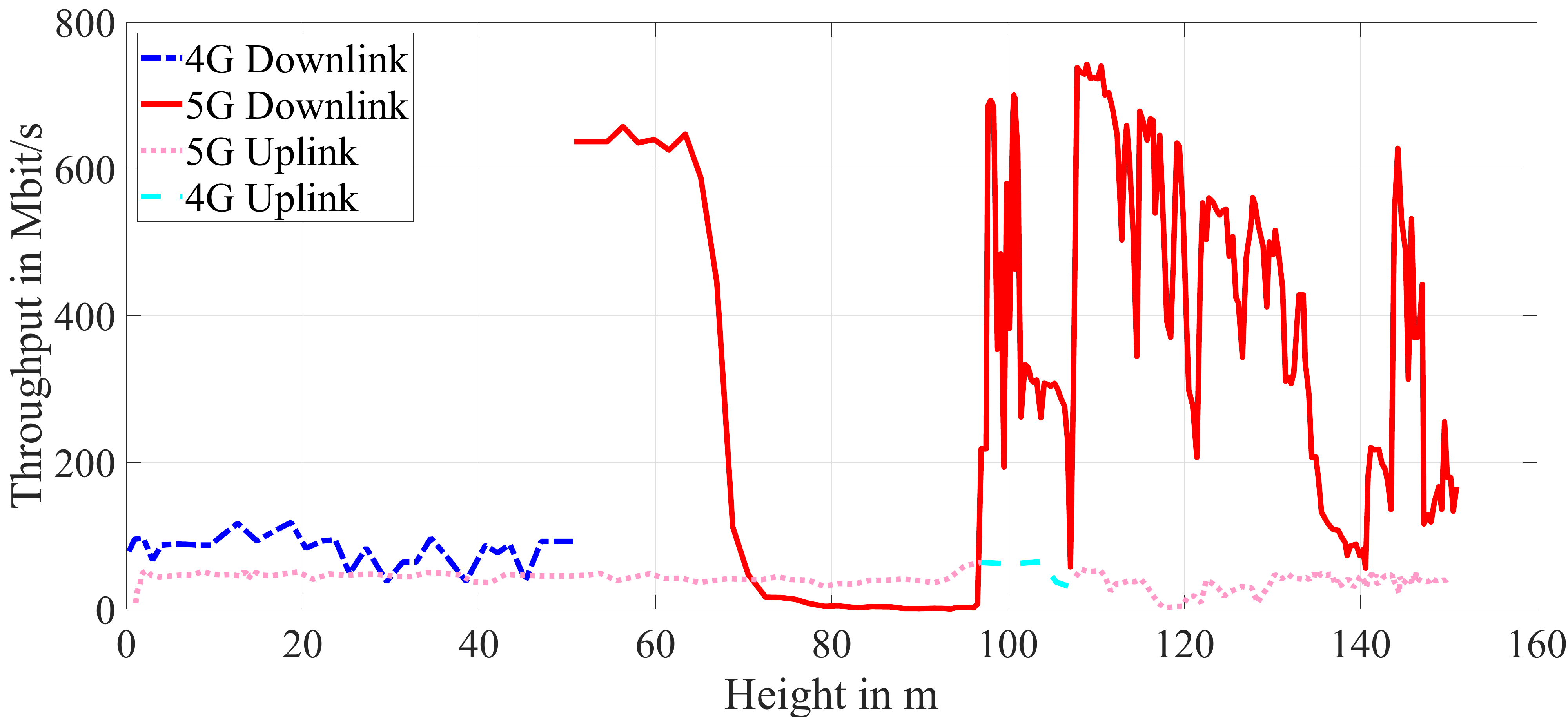} \\
\hspace{-0.3cm}(c) Throughput\\[0.1cm]
\renewcommand{\arraystretch}{1}
\end{center}
\caption{Liftoff phase: Radio link performance of a drone flying from the ground to a height of 150\,m}
\label{fig:Vertical_DL_UL}
\end{figure}

Fig.~\ref{fig:Vertical_DL_UL} shows the RSRP, SNR, and throughput during a liftoff from the ground to a height of $150$~m. In the DL, the drone initially connects to 4G  but performs a handover to 5G at a height of $50$~m. A similar behavior occurs in the UL: the drone initially connects to 5G, switches to 4G at a height of $97$~m but  switches back to 5G at $107$~m. The DL throughput over 5G is $387$~Mbit/s on average but shows high fluctuations with peaks above $700$~Mbit/s. Although the antennas reconfigure electrically for beamforming, the high  fluctuation could be caused by connectivity with the side lobes (which have lower radiation intensity than the main lobe). The DL throughput in 4G is $83$~Mbit/s on average with peaks up to $118$~Mbit/s. The average UL throughput is $53$~Mbit/s in 4G and only $39$~Mbit/s in 5G. This discrepancy requires further investigation. 

\begin{figure}[t!]
\begin{center}
\centering 
\includegraphics[width=3.4in]{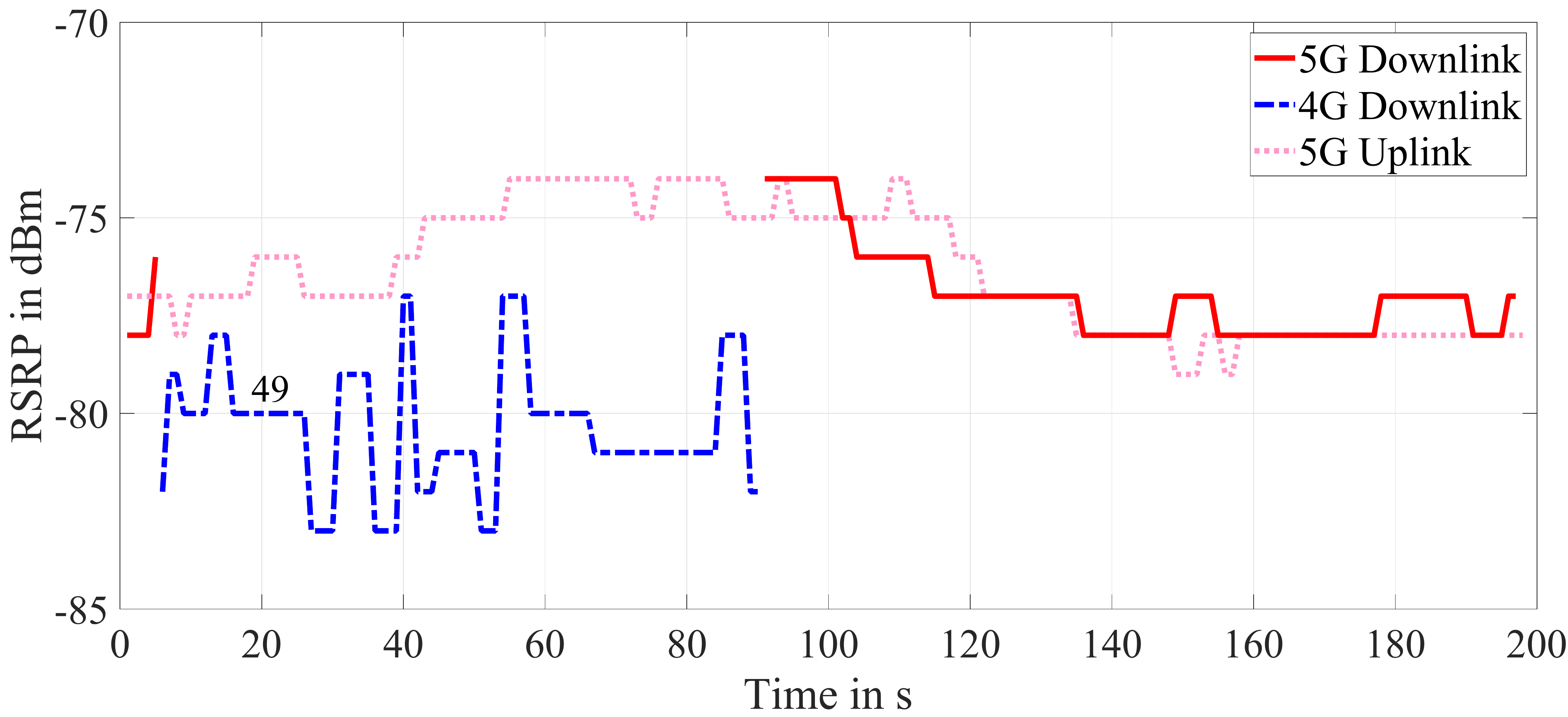} \\ 
\hspace{-0.3cm}(a) RSRP \\[0.3cm]
\includegraphics[width=3.4in]{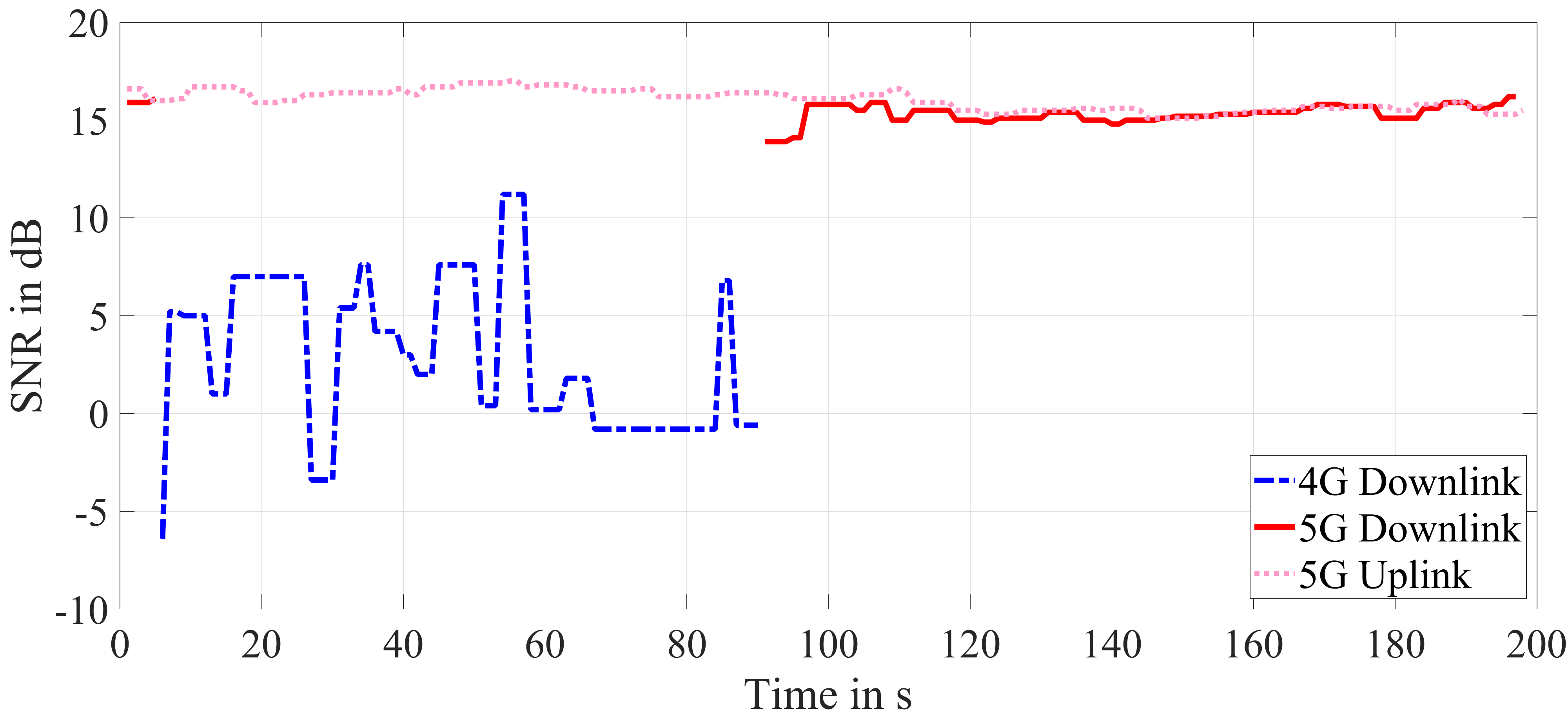} \\
\hspace{-0.3cm}(b) SNR \\[0.3cm]
\includegraphics[width=3.4in]{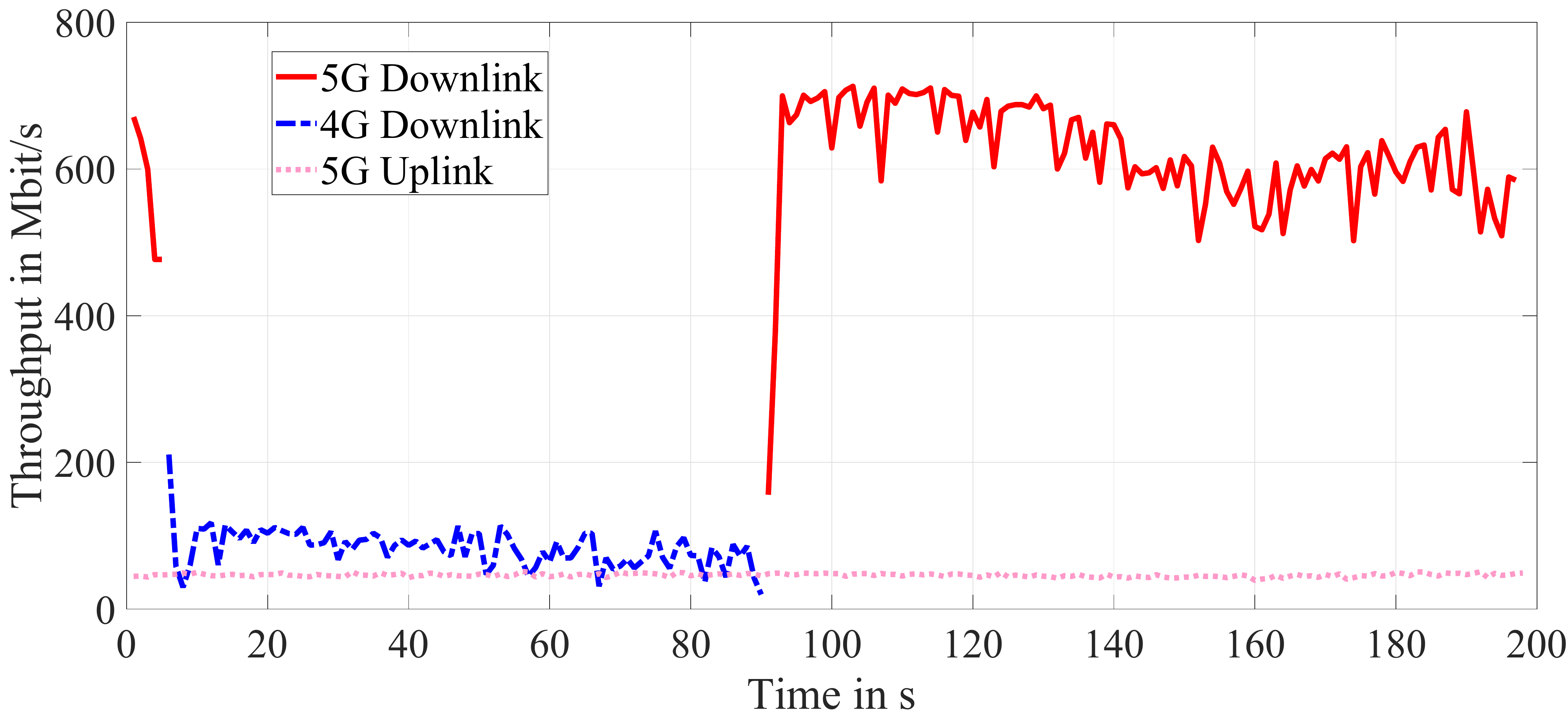} \\
\hspace{-0.3cm}(c) Throughput\\[0.1cm]
\renewcommand{\arraystretch}{1}
\end{center}
\caption{Horizontal flight~1: Radio link performance of a drone flying at 30~m height}
\label{fig:30m_DL_UL}
\end{figure}

\subsection{Horizontal flights}

Fig.~\ref{fig:30m_DL_UL} shows the RSRP, SNR, and throughput for a horizontal flight at $30$~m height. In the DL, the drone first connects to 5G but immediately switches to 4G due to a disconnection to~5G. As soon as the 5G connectivity is reestablished, a handover back to 5G occurs. Interestingly, such handovers do not occur in the UL: the drone always remains connected to 5G with an average SNR of about $16$~dB. 
Similar to the results during the liftoff, the DL throughput is much higher in 5G than in 4G, namely $618$~Mbit/s on average with peaks above $700$~Mbit/s, compared to an average 4G throughput of $83$~Mbit/s with peaks of about $200$~Mbit/s. However, compared to the liftoff, the fluctuations of the throughput are less severe:  the standard deviation is $77$~Mbit/s compared to more than $200$~Mbit/s for the liftoff. In the UL, the throughput remains stable with an average of $46$~Mbit/s and a standard deviation of $2$~Mbit/s.

\begin{figure*}[t!]
\begin{center}
\centering 
\textbf{Downlink} \quad \quad \quad \quad \quad \quad \quad\quad \quad\quad \quad\quad \quad\quad \quad\quad\quad\quad \quad\quad \quad\quad \textbf{Uplink}\\[0.3cm]
\includegraphics[width=3.4in]{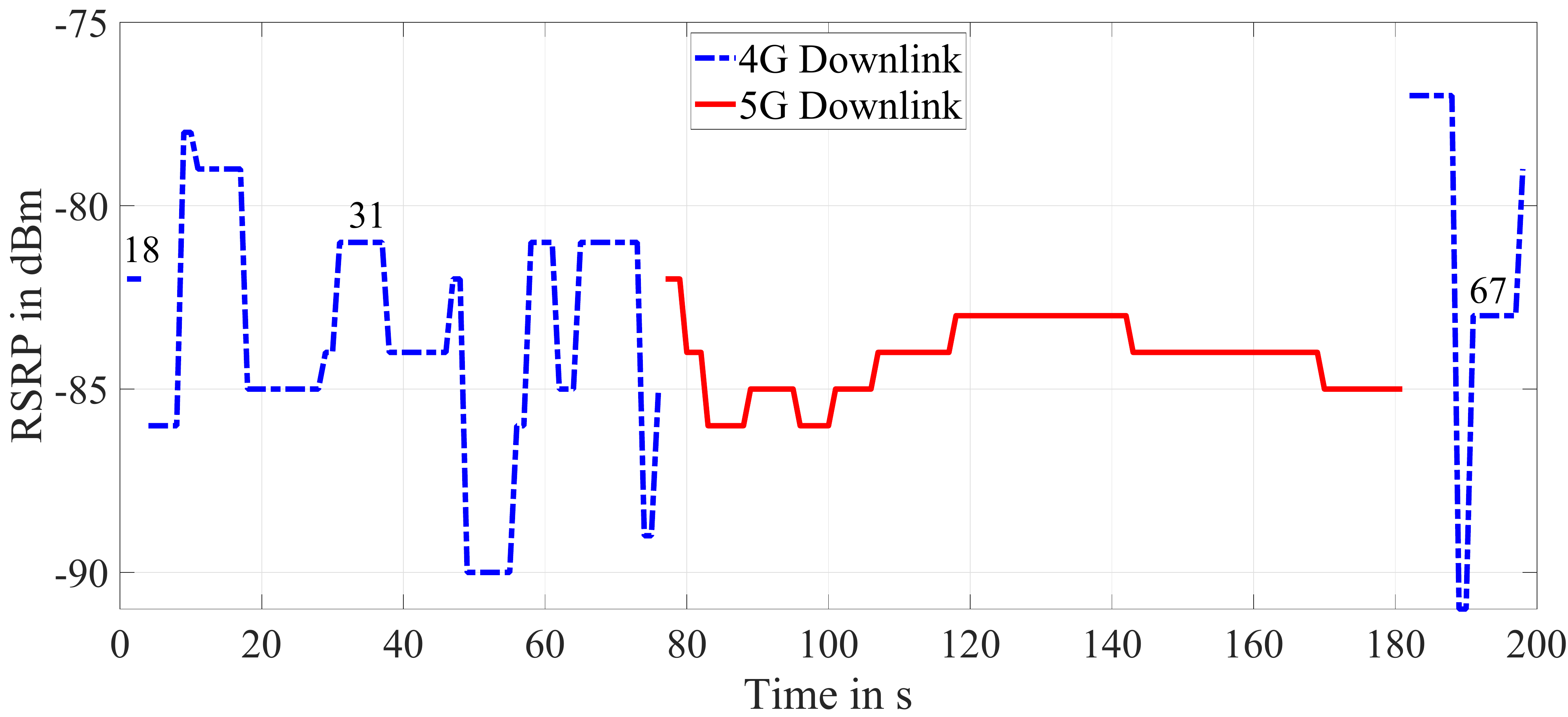}~~~~\includegraphics[width=3.4in]{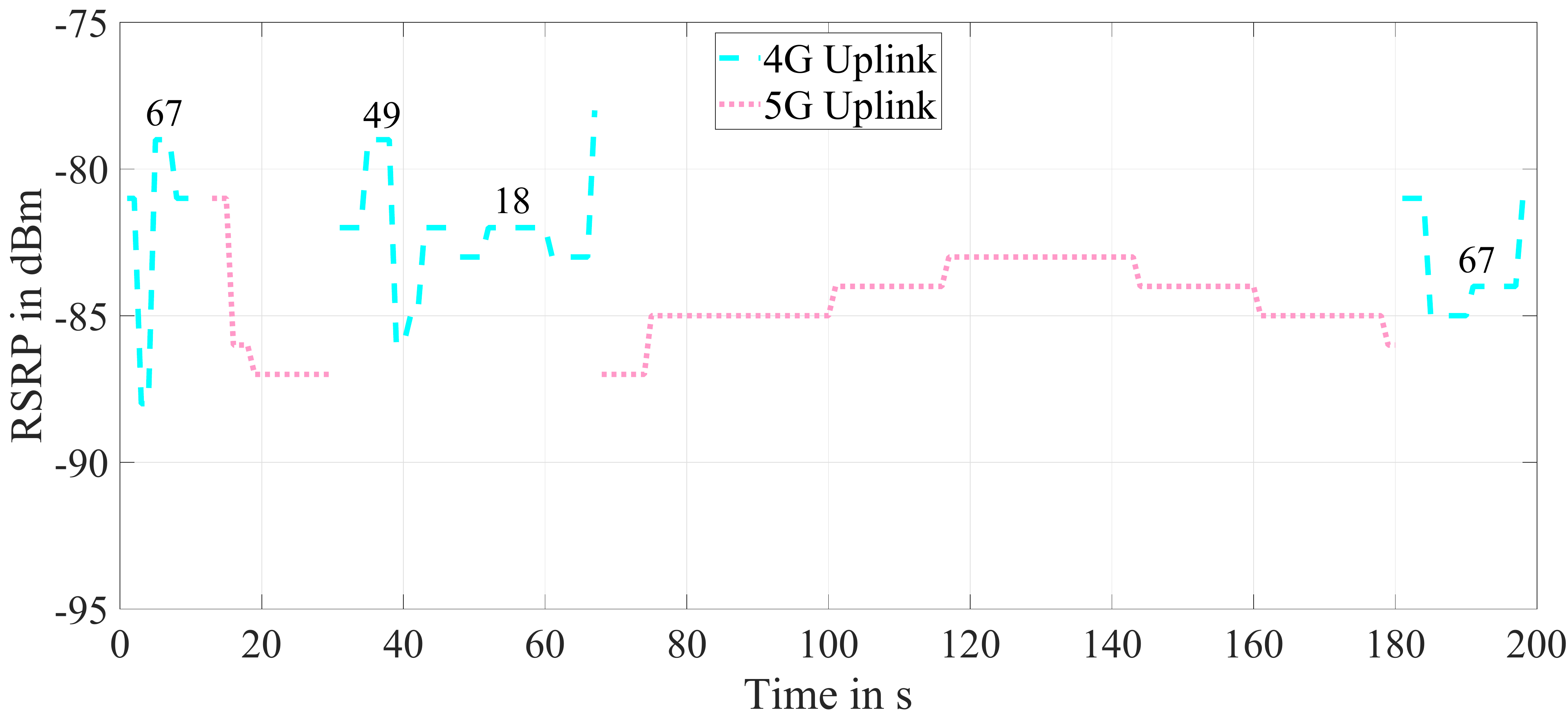}\\
\hspace{-0.3cm}(a) RSRP \\[0.3cm]
\includegraphics[width=3.4in]{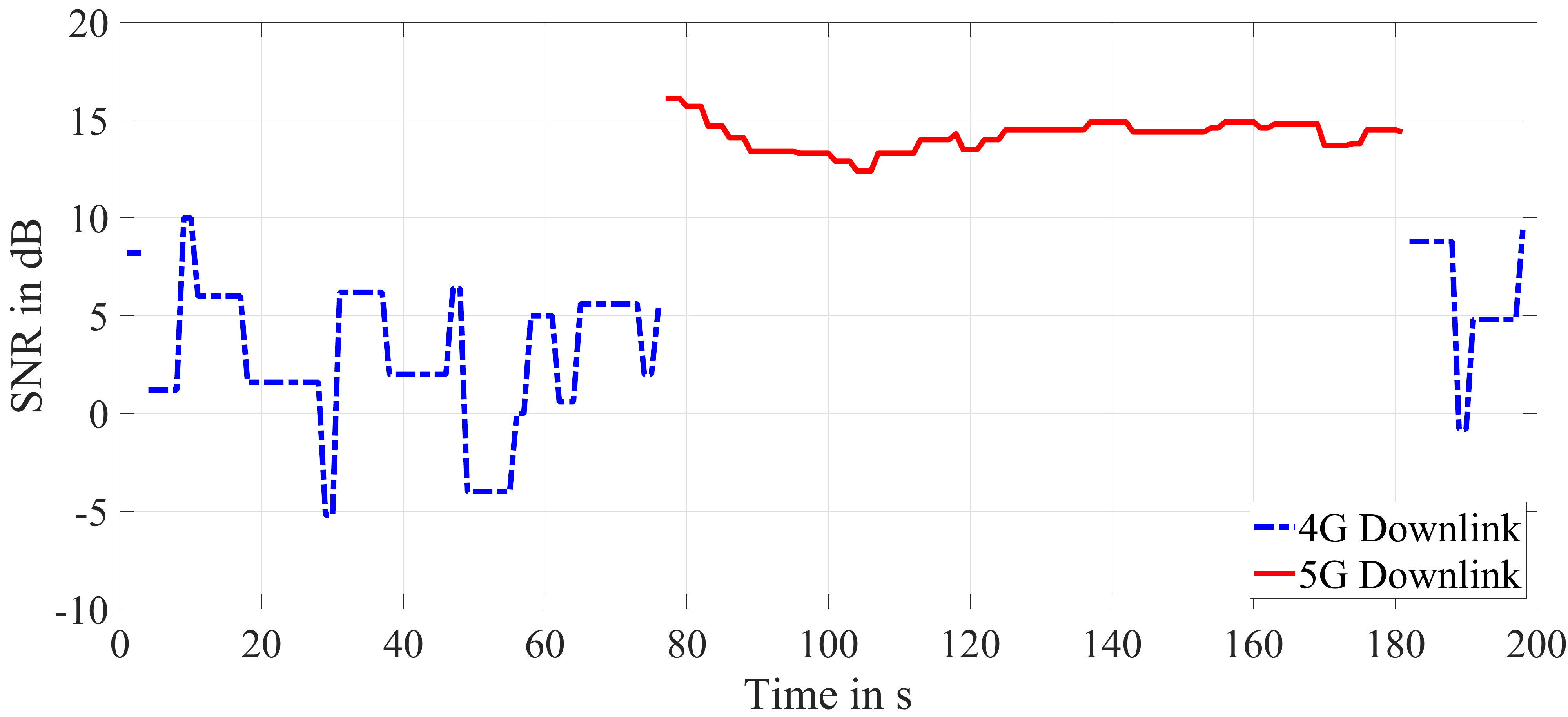}~~~~\includegraphics[width=3.4in]{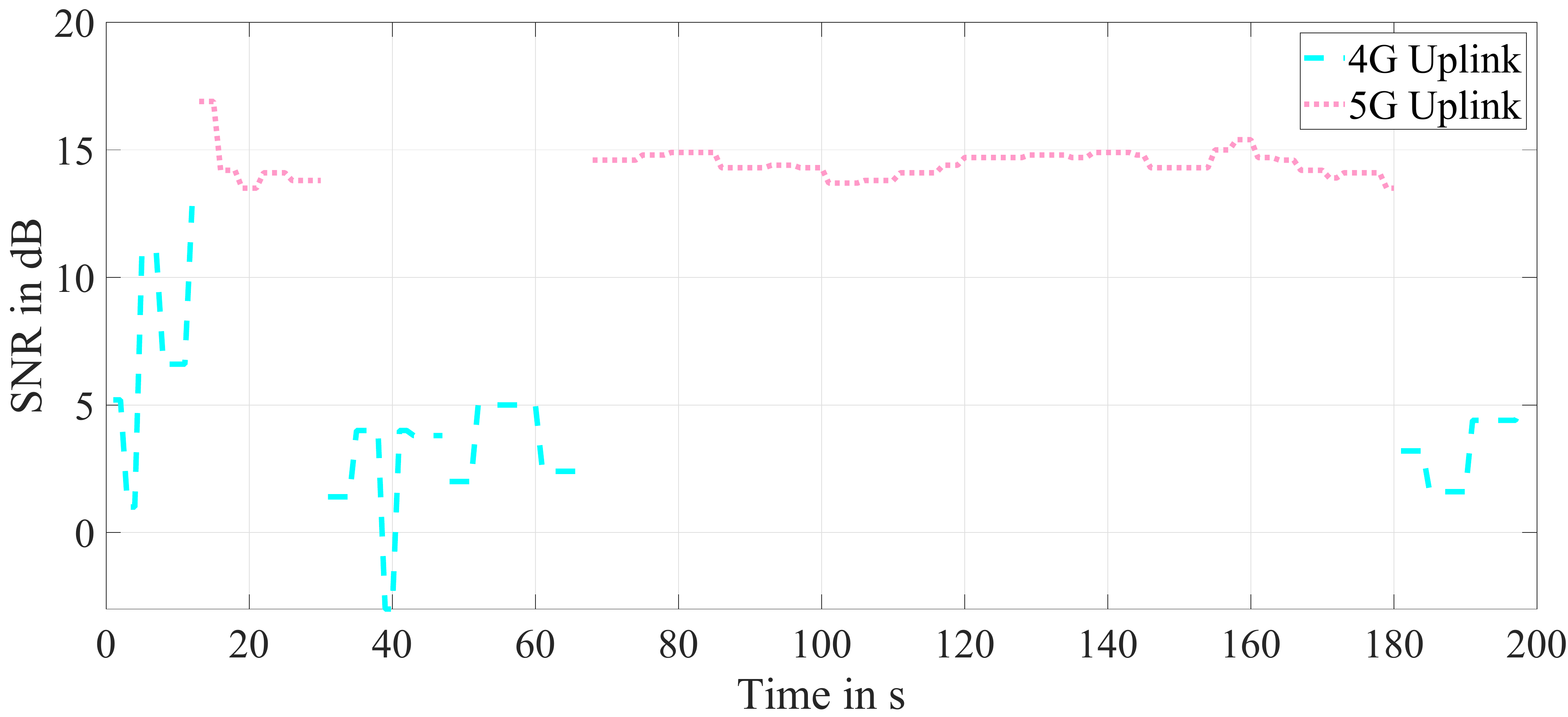} \\
\hspace{-0.3cm}(b) SNR \\[0.3cm]
\includegraphics[width=3.4in]{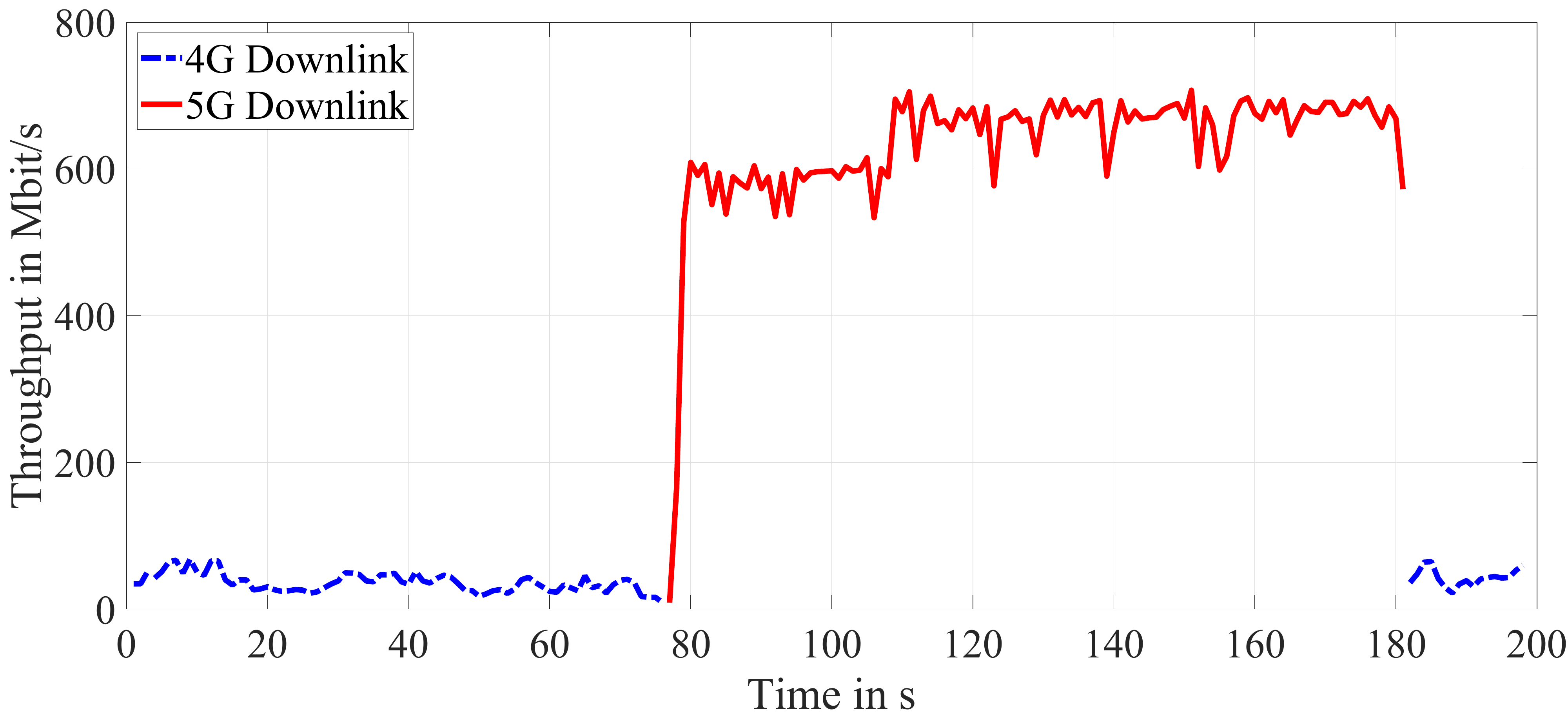}~~~~\includegraphics[width=3.4in]{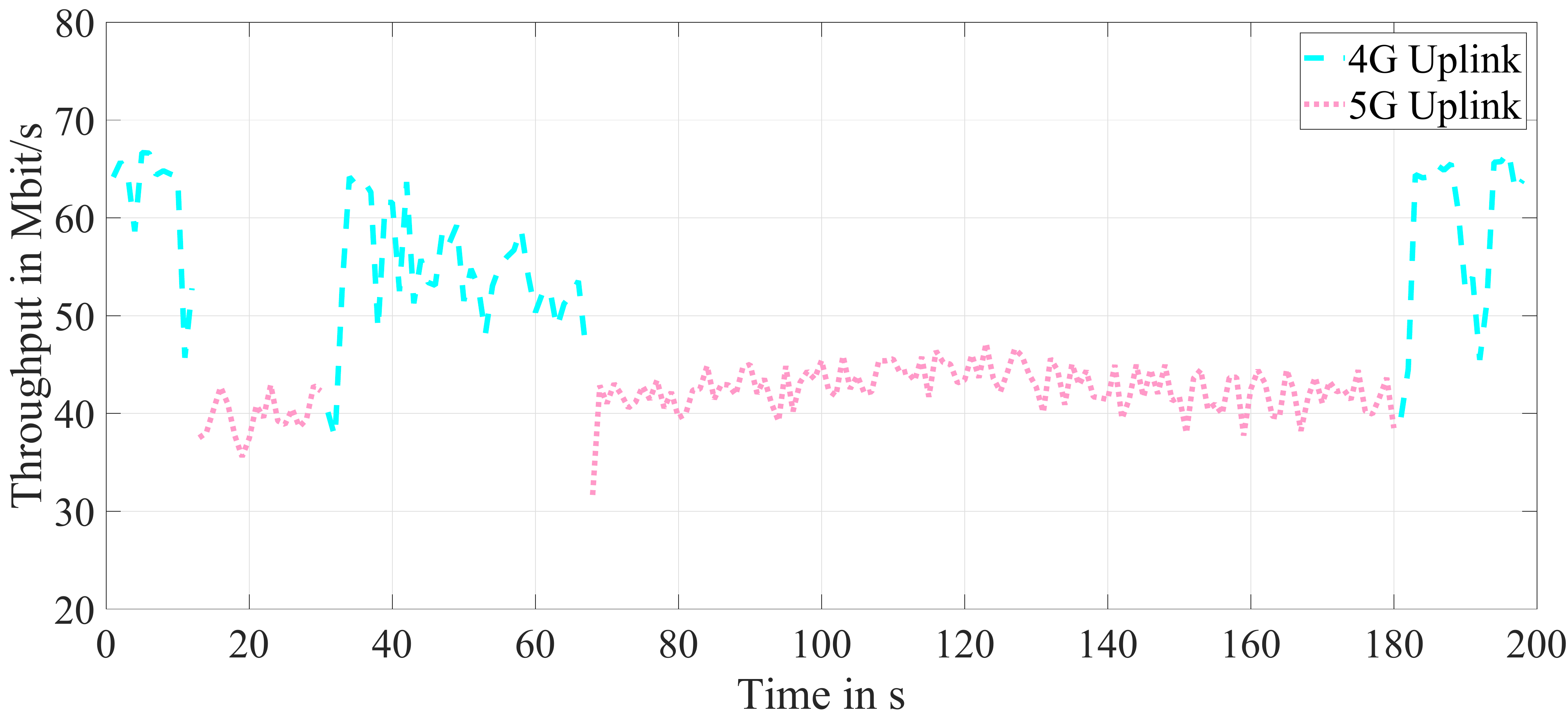} \\
\hspace{-0.3cm}(c) Throughput\\[0cm]
\renewcommand{\arraystretch}{1}
\end{center}
\caption{Horizontal flight~2: Radio link performance of a drone flying at 100~m}
\label{fig:100m_DL}\label{fig:100m_UL}
\end{figure*}

Finally, Fig.~\ref{fig:100m_DL} shows the results when flying at $100$~m. The increased height leads to more handovers (DL:~3; UL:~5). This is most likely because the drone connects to distant 4G BSs via side lobes that have better signal quality than the physically closest BS (as in~\cite{fakhreddine19}). The average DL throughput is $644$~Mbit/s for 5G and $37$~Mbit/s for 4G. The average UL throughput is again lower for 5G ($42$~Mbit/s) than for~4G~($56$~Mbit/s).

\section{Conclusions and Outlook}\label{concl}

We performed radio link measurements with a drone in a region with a 5G base station that is otherwise covered by 4G. The link quality, throughput, and handovers were analyzed for the liftoff and horizontal flights at two typical~heights. 

Some specific quantitative results can be generalized as~follows. First, the connectivity to the 5G base station cannot be maintained during the entire flight but handovers to 4G occur. As the flight height increases, the throughput suffers and handovers become more frequent. 

Second, drones at typical flight heights can receive several hundred Mbit/s over 5G, which is sufficient for many applications. However, we have to keep in mind that the 5G DL performance is biased by the fact that there is only one 5G BS. Broader deployment of 5G will change the setup with unclear consequences. On the one hand, more 5G BSs on the same frequency are expected to cause interference to airborne drones, due to line-of-sight links, which would lower the 5G throughput (as discussed for 4G in \cite{van2016lte}). On the other hand, further 5G BSs will diminish handovers to 4G and thus in turn improve the overall DL throughput. 

Third, the 5G implementation used does not improve the UL throughput compared to 4G. This issue needs further investigation, especially as the UL is important for drones to send pictures and videos to~the~BS.

\balance
\section*{Acknowledgements}
This work has been performed in a collaboration between the University of Klagenfurt, Lakeside Labs GmbH, and Magenta Telekom GmbH (previously T-Mobile Austria GmbH). R.~Muzaffar and C.~Raffelsberger were funded by the security research program KIRAS of the Federal Ministry for Transport, Innovation, and Technology (BMVIT), Austria, under grant agreement n.~854747 (WatchDog). C.~Bettstetter is faculty member in the Karl Popper Kolleg on networked autonomous aerial vehicles at the University of Klagenfurt. 
\bibliographystyle{unsrt}
{
\bibliography{Ref}}
\balance
\end{document}